# Quantum yield of an emitter proximate to nanostructures: a quantitative understanding

Jinsong Duan [1,2]*

1  Department of Physics, University of Dayton, 300 College Park, Dayton, OH 45469

2  General Simulation, LLC, 1902 Founders Dr, Dayton, OH 45420

*  Correspondence: jinsduan@gmail.com

**Abstract:** Exciton-surface plasmon coupling is at the heart of the most elementary light-matter interactions and is a result of not only an intrinsic property of the emitter but that of emitter-environment interaction. Thus, change of electromagnetic environment, as in case of metallic nanoplasmonic structures and an emitter, significantly modifies the near field light-matter interaction, which leads to energy transfer in the form of exciton between metallic nanostructure and the emitter. However, this mechanism remains largely unexplored. Here, we developed and applied semi-classical electrodynamics theory and modeling techniques to analyze the energy transfer mechanism in exciton-surface plasmon coupling. The quantum efficiency of an emitter was investigated as a function of the location of the emitter with respect to nanoparticles and their assembles whose local plasmonic field modified by forming complex coupling modes as well as the local dielectric environment. The research provided a theoretical insight into fundamental science of nanophotonics and shed light on unprecedented applications in wide range fields such as ultra-low power lasers, quantum information processing, photovoltaics, photocatalysis, and chemical sensing.

## 1. Introduction

The last decades have seen a rapid demand in global energy use. Renewable energy sources have been becoming more attractive than the traditional energy sources which have negative environmental impacts such as climate change and global warming. Among all the renewable energy sources, solar power, which is abundant, free, and clean, in particular, has attracted more attentions as the greatest promising option to solve increasing energy demand without bringing any pollution to the environment. [1,2]

The most promising renewable energy technologies are solar photovoltaics and photocatalysis, which convert solar energy into electricity and chemical energy, respectively.[3,4] Realizing more efficient solar energy conversion, which is a key driver to reduce the cost of solar energy, depends on the light absorption properties, i.e., enhanced light coupling and trapping, of photovoltaic and photocatalytic devices. [5,6]

Light emission and absorption are reprocial physical processes. One expects a reciprocal law governing two processes, as proposed by Rau. [7] That is an absorber has the theoretical maximum power conversation efficiency will also act as an emitter with the maximum possible luminescence efficiency. Therefore, understanding the enhancement of light emission process is crucial to enhance light coupling and trapping in photovoltaics and photocatalysis devices.

In 1946, Purcell discovered that a quantum system's spontaneous rate by can be enhanced by its environment. [8]

The spontaneous emission of am emitter in free space is,

$$\Gamma_0 = \frac{\omega_0^3 \mu_{12}^2}{3\pi\epsilon_0 \hbar c^3} \tag{1}$$

where $\mu_{12}^2$ is the transition dipole matrix between the excited state and ground state.

The modified emission rate of an emitter in non-free space, such as in the presence of a cavity, is defined as,

$$\Gamma_g = \frac{2\pi \mu_{12}^2 E_0^2}{\hbar^2} \rho(\omega_0) \tag{2}$$

where $E_0$ is the electric field created by the emitter and $\rho(\omega_0)$ is the local density state. The Purcell effect is the enhancement of a quantum system's spontaneous emission rate by its environment and can be characterized by the Purcell factor, which is the ratio between equation 1 and 2,

$$F_p = \frac{\Gamma_g}{\Gamma_0} = \frac{3Q\lambda^3}{4\pi^2 V_0} \tag{3}$$

where $Q$ is the quality factor of a cavity and $V_0$ is its volume.

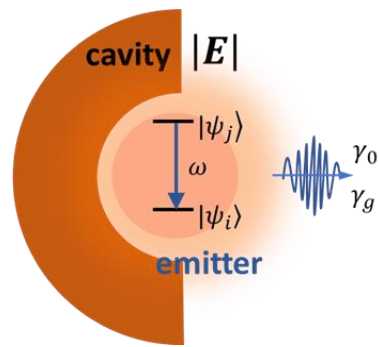

**Figure 1.** A quantum emitter proximate to a cavity. Its intrinsic quantum yield $\gamma_0$ is modified to be $\gamma_g$ due to the prensence of the cavity through Purcell effect.

Equation 3 provides a very useful insight that the spontaneous emission of an emitter can be enhanced by confining the light into a small volume by an optical cavity.

Metallic nanoparticles can spatially confine light on a nanometer scale by trapping energy into localized surface plasmon modes,[9] which depends on material and geometries as well as the local dielectric environment, [10–14], can be used as optical cavities to improve the quantum efficiency of an emitter. [15]

Semiconductor materials have demonstrated interesting optoelctronic properties at optical and infrared range and are attractive materials in harvesting solar energy. [16,17] As such hybrid metallic nanoparticles-semiconductor architectures offer promising platforms for high-efficient light harvesting, where light absorption in semiconductor materials can be enhanced in the presence of metallic nanoparticles through photon-exciton coupling mechanisms. However, the photon-exciton coupling mechanism is largely unknown and need to be quantitatively understood.

In this work, we systematically study the quantum efficiency of an emitter near metallic particles and their nanostructures focusing on the distance, orientation, geometries of nanostructures.

## 2. Materials and Methods

Gold nanoparticles has the capability to confine light into small volume due to the excitation of surface plamonics. The three dimensional electric field around a rod-like gold nanorod is shown in Figure 2.

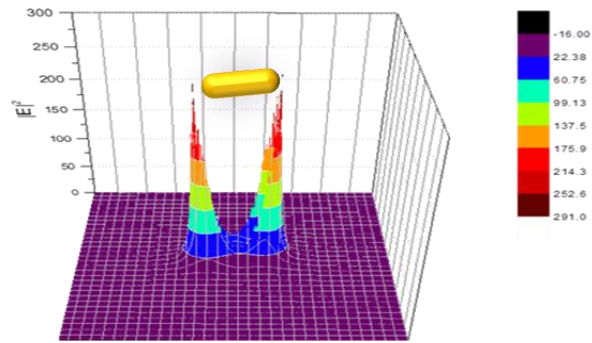

**Figure 2.** The 3-dimensional electric field distribution around a gold nanorod.

The magnitude of the electric field increases exponentially as the distance to the surface of the gold nanorod decreases, in particular at the tips of the rod. the electric field enhancement is defined as the ratio between the integral of the electric field around the rod and that of the incident light,

$$F = \frac{\iiint_v |E|^2 dx dy dz}{\iiint_v |E_0|^2 dx dy dz} \tag{4}$$

Plasmonic enhanced electric field makes plasmonic nanostructures good candidate for optical cavity to modify the quantum yield of an emitter based on the Purcell effect. To quantitatively understanding the quantum efficiency of an emitter close to nanostructures, we need to analyze the energy transfer between the emitter and nanosturctures.

Once plasmons in the nanostructures are excited by the emitter, the plasmons can radiate into free space as radiative decay, at rate $r_r^{pl}$, or they can be lost to absorption as non-radiative decay, at rate $r_{nr}^{pl}$. At the same time, the emitter radiate to the free space at the rate $r_r'$. The details is schematically illustrated in Figure 3. The total decay rate, $\Gamma_{tot}$, can be written as,

$$\Gamma_{tot} = \gamma_r' + \gamma_{nr}^{pl} + \gamma_r^{pl} \tag{5}$$

where $\gamma_{nr}^{pl} + \gamma_r^{pl}$ is plasmon decay in the nanostructure excited by the emitter in the form of the energy transfer.

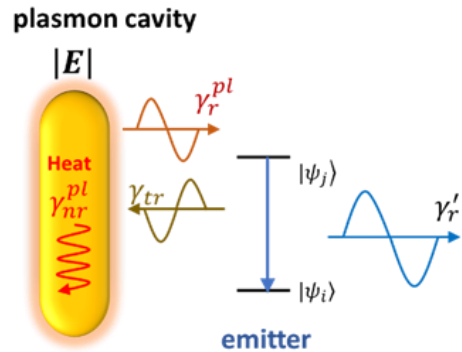

**Figure 3.** Schematical energy conservation in a system consisting of an emitter and a nanorod. The emitter radiates to the free space at the rate of $\gamma'_r$ and transfers energy to the nanostructure at the rate of $\gamma_{tr}$, which splits into the plasmonic radiation $\gamma_r^{pl}$, and the plasmonic loss $\gamma_{nr}^{pl}$ in form of heat.

The radiative quantum yield of an emitter in the absence of metallic nanoparticles is written as,

$$Y = \frac{\gamma_0}{(\gamma_0 + \gamma_{nr})} \quad (6)$$

however, in the presence of metallic nanoparticle, the quantum efficiency of an emitter becomes,

$$Y = \frac{\gamma' + \gamma_r^{pl}}{\gamma' + \gamma_{tr}} = \frac{\gamma_r}{\gamma_r + \gamma_{nr}} \quad (7)$$

where $\gamma' + \gamma_r^{pl}$ is the radiative decay rate $\gamma_r$, $\gamma' + \gamma_{tr}$ is the total decay rate, i.e., the summation of the radiative decay rate $\gamma_r$ and non-radiative decay rate $\gamma_{nr}$, both can be calculated numerically as discussed below.

To elaborate the detailed information about the energy transfer, the quantum yield can be written as,

$$Y = \frac{\gamma'_r}{\gamma'_r + \gamma_{tr}} + \frac{\gamma_{tr}}{\gamma' + \gamma_{tr}} \cdot \frac{\gamma_r^{pl}}{\gamma_r^{pl} + \gamma_{nr}^{pl}} \quad (8)$$

where the second term addresses the energy transfer between the emitter and nanoplasmonic structure, $\gamma_{tr} = \gamma_r^{pl} + \gamma_{nr}^{pl}$.

It is different to calculate Purcell factor directly from Purcell effect. However, it can be calculated using the energy conservation, as stated in Poynting theorem, that energy flux of propagating electomagentic wave is conserved.

Consider a system consists of a nanorod with a quantum emitter next to it as shown in the Figure 4, the energy radiates out of the simulation region equals to the energy emits from the emitter minus the energy loss as heat in the nanorod. That is,

$$\iint_S (E \times H) ds + \frac{1}{2} \frac{\partial}{\partial t} \iiint_v (\epsilon E^2 + \mu H^2) dv + \iiint_v (E \cdot J) dv = 0, \qquad (9)$$

where the first term is the integral of energy radiating out of the simulation volume, the second term is energy stored in the simulation volume, the third term is the energy loss.

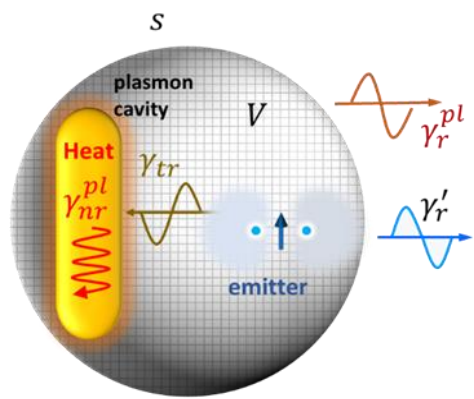

**Figure 4.** The simulation setup for calculating the energy between a quantum emitter and a nanorod. Radiative decay, $\gamma_r$, is the summation of $\gamma_r^{pl}$ and $\gamma'$.

radiative decay rate, $\gamma_r$:

$$\gamma_r = \frac{\gamma_r^0}{W_r^0} \iint_S \frac{1}{2} Re(E \times H) ds \qquad (10)$$

non-radiative decay rate, $\gamma_{nr}$:

$$\gamma_{nr} = \frac{\gamma_r^0}{W_r^0} \iiint_v \frac{1}{2} Re(J \cdot E) dv \qquad (11)$$

## 3. Results

### 3.1. Distance between an emitter and nanoparticle

The quantum efficiency of an emitter demonstrates a bi-model feature as a function of its distance from a nanorod, as shown in Figure 5. That is, as the emitter moves closer to the nanorod, both the radiative and non-radiative decay increase with the non-radiative decay quickly becoming dominative when the distance is smaller than 2 nm, making the quantum efficiency decreasing dramatically. At a very short distance, i.e., less than 2 nm, plasmonic loss plays a major role in energy balance, reducing the radiative energy term.

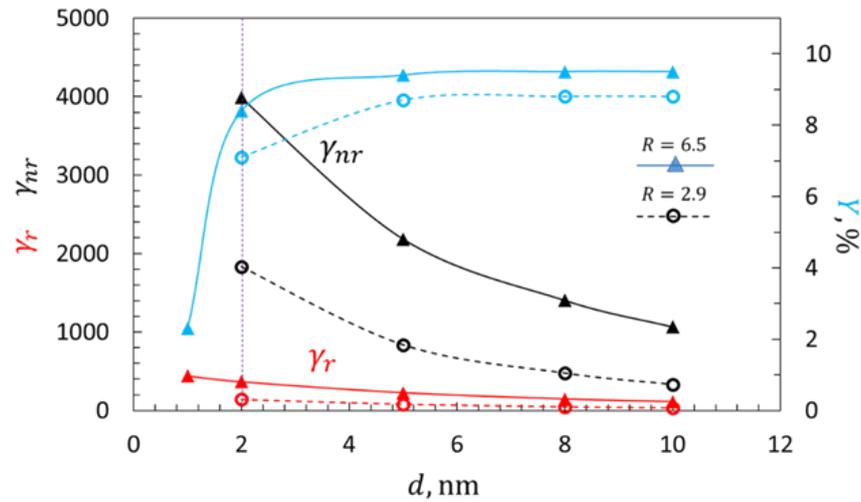

**Figure 5.** Quantum efficiency $Y$, radiative $\gamma_r$ and non-raditive $\gamma_{nr}$ decay of an emitter as a function of the distance from the nanorod.

*3.2. Nanoparticle sizes*

The lower energy resonance mode is excited in the longer nanorod. [14] A lower energy mode reduces the plamonic energy loss in the rod, increasing the radiative energy. It is demonstrated exactly in Figure 5, where the quantum efficiency of a longer nanorod with the aspect ratio of 6.5 is higher than that of the rod of 2.6 aspect ratio.

*3.3. Nanoparticle assemble*

The quantum efficiency of an emitter can also be turned by coupling variety of nanostructures. We consider a nanorod pair separated by 5 nm with a emitter located in the middle with its polarization parallel to the long axes of the rods, as illustrated in Figure 6. The emitter has the quantum yield of 10.5%, compared with that of 5% for an emitter at the end of a single rod. Anti-bonding mode of high energy is excited with the emitter, providing the stronger radiative decay overtakes the plasmonic loss, which can be also significant, considering the distance to the emitter is 2.5 nm.

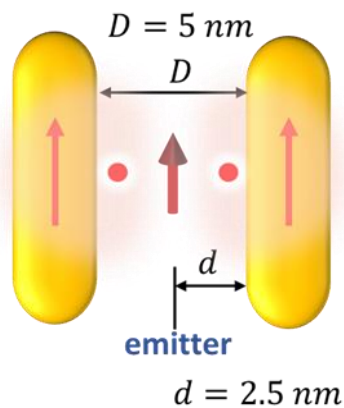

**Figure 6.** An emitter located in the gap between two nanorods with its polarization parallel to the long axes of the nanorod pair.

*3.4 Local dielectric environment*

Understanding the quantum yield of an emitter near nanoparticle in various media has important applications in design ultra-sensitive biosensors. As a matter of fact, the plasmonic resonance mode in nanoparticle depends on the local dielectric environment. [14] The lower energy mode is excited in dielectric environment than in the vacuum. In the presence of aqueous environment, for instance, the quantum yield of an emitter (R=2.9) separated from the tip of a nanorod by 10 nm is increased to 13.7 %, compared to 8.8 % in the same configure, while in the vacuum.

**5. Conclusions**

In summary, we systematically analysis and quantify the quantum yield of an emitter proximate to metallic plasmonic nanostructures which are promising candidates to enhance light-matter interaction by confining light in small volume. The quantum yield of an emitter is quantified by its size, position, orientation, and dielectric environment with respect to nanostructures.